\documentstyle[epsf]{mn}
\title[Stellar dynamics\ldots]{Stellar dynamics in a galactic centre
 surrounded by a massive accretion disc. I.~Newtonian description}
\author[D.~Vokrouhlick\'y and
 V.~Karas]{D.~Vokrouhlick\'y$^{\,1,2\,}$\thanks{E-mail: 
 vokrouhl@mbox.cesnet.cz} and
 V. Karas$^{\,1,3\,}$\thanks{E-mail: karas@mbox.troja.mff.cuni.cz}\\
 $^{1}$Astronomical Institute, Charles University Prague,
 V~Hole\v{s}ovi\v{c}k\'ach~2, CZ-180\,00 Praha, Czech Republic\\
 $^2$Observatoire de la C\^ote d'Azur, dept.~CERGA, Av.~N.~Copernic,
  F-06130~Grasse, France\\
 $^3$Scuola Internazionale Superiore di Studi Avanzati, Via~Beirut~4,
 I-34014~Trieste, Italy}
\date{Received \today}
\pagerange{\pageref{firstpage}--\pageref{lastpage}}

\def\spose#1{\hbox to 0pt{#1\hss}}
\def\lta{\mathrel{\spose{\lower 3pt\hbox{$\mathchar"218$}}
     \raise 2.0pt\hbox{$\mathchar"13C$}}}
\def\gta{\mathrel{\spose{\lower 3pt\hbox{$\mathchar"218$}}
     \raise 2.0pt\hbox{$\mathchar"13E$}}}
\newcommand{\myder}{{\mbox{d}}}
\newcommand{\beq}{\begin{equation}}
\newcommand{\eeq}{\end{equation}}
\newcommand{\di}{{\rm{}d}}
\newcommand{\uni}{{\rm{}u}}
\newcommand{\BH}{{\rm{}c}}
\newcommand{\const}{{\rm{}const}}
\newcommand{\etal}{{\rm{}et~al.\ }}
\def\lb#1{{\protect\linebreak[#1]}}
\def\hp{{http://\lb{2}astro.troja.\lb{2}mff.cuni.cz/}}

\font\tenmsbm=msbm10
\font\sevenmsbm=msbm7
\font\fivemsbm=msbm5
\newfam\msbmfam
\textfont\msbmfam=\tenmsbm
\scriptfont\msbmfam=\sevenmsbm
\scriptscriptfont\msbmfam=\fivemsbm
\def\akpa{\mathchar"2B7B} 

\begin{document}
\label{firstpage}
\maketitle

\begin{abstract}
 The long-term evolution of stellar orbits bound to a massive centre is
 studied in order to understand the cores of star clusters in central
 regions of galaxies. Stellar trajectories undergo tiny perturbation,
 the origin of which is twofold: (i)~gravitational field of a thin gaseous
 disc surrounding the galactic centre, and (ii)~cumulative drag due to
 successive interactions of the stars with material of the disc. Both
 effects are closely related because they depend on the total mass of
 the disc, assumed to be a small fraction of the central mass. It is
 shown that, in contrast to previous works, most of the retrograde (with
 respect to the disc) orbits are captured by the central object,
 presumably a massive black hole. Initially prograde orbits are also
 affected, so that statistical properties of the central star cluster in
 quasi-equilibrium may differ significantly from those deduced in
 previous analyses.

\end{abstract}


\section{Introduction}
This paper extends previous studies on interaction between stars and an
accretion disc near a massive galactic nucleus. Relevant references are,
in particular, Syer, Clarke \& Rees (1991, these authors estimate
time-scales for the evolution of stellar orbital parameters in the
Newtonian regime), and Vokrouhlick\'y \& Karas (1993, relativistic
generalization dealing with individual trajectories). Pineault \& Landry
(1994) and Rauch (1995) studied statistical properties of stellar orbits
in a dense cluster near a galactic core with an accretion disc.

Observational evidence and theoretical considerations suggest that many
galaxies harbour very massive compact cores
($M_\BH\approx10^6$--$10^9M_\odot$), presumably black holes. In
particular, high energy output, variability, spectral properties, and
production of jets in active galactic nuclei (AGN) can be understood in
terms of the model with a supermassive central object surrounded by
an accretion disc (e.g., Courvoisier \& Mayor 1990; Urry \& Padovani
1995). However, linear resolution of present observational techniques
corresponds at best to several hundreds of gravitational radii of the
hypothetical black hole. The innermost regions of these galaxies cannot
thus be resolved and conclusions about their structure must be inferred
from integral characteristics (integrated over the angular and temporal
resolution of the device used for observation). Distribution of stars
and gaseous material close to the galactic centre is one of the important
tools in this respect because velocity dispersion and the corresponding
luminosity profile of the nucleus reflect the presence and properties of
the central massive object and the disc (Perry \& Williams 1993; cf.\
Marconi et al.\ 1997 for recent observational results).

We will study the situation in which the central object is surrounded by an
accretion disc and a dense star cluster. It is the aim of the present
contribution to examine the role of periodic interactions of the stars
with the disc material, {\em{}simultaneously\/} considering the gravitation
of the disc. Mutual gravitational interaction of stars forming a dense
cluster has been studied since the early works of Ambartsumian (1938)
and Spitzer (1940) while the importance of star-disc collisions for the
structure of galactic nuclei has been recognized in the early 1980s
(Goldreich \& Tremaine 1980; Ostriker 1983; Hagio 1987). Huang \&
Carlberg (1997) studied a related problem in the dynamics of galaxies.

The gravitation of accretion discs was neglected in previous works because
its mass, $M_\di$, is presumed very small compared to the mass of the
central object ($\mu\equiv{}M_\di/M_\BH\ll1$; $\mu$ is a free parameter in our 
study). We also assume $\mu\ll1$ so that the gravity of the
disc acts as a perturbation on the stellar motion around the central mass.
We will show, independent of the precise value of $\mu$, 
that the effect of the disc gravity on the
long-term evolution of stellar orbits must be taken into account
together with star-disc interactions. In particular, we will show that
circularization of many of the orbits, evolution of their inclination, and
stellar capture rate are visibly affected by the disc gravity. We
will also argue that the physical reason for this fact is the existence of
three different time-scales involved in the problem: (i)~the orbital
period of the star around the central mass (short time-scale),
(ii)~the period of oscillations in eccentricity and inclination of the orbit
(medium time-scale, these oscillations are due to the disc gravity), and
(iii)~the time for grinding the orbital plane into the disc (long
time-scale, due to successive interactions of the stars with the disc).
Effects which can be ascribed to the medium time-scale present a new
feature discussed in this paper within the context of galactic nuclei
surrounded by an accretion disc, although analogous effects of
oscillations or sudden changes in orbital parameters are well-known from
other applications (cf.\ recent discussion on dynamics of planetary
motion by Holman, Touma \& Tremaine 1997; Lin \& Ida 1997).

Details of the model are described in the next section. Then, in
Sec.~\ref{evolution}, long-term evolution of stellar orbits is examined.
Finally, conclusions of our present paper are summarized in
Sec.~\ref{conclusions}.

\section{The model}
\label{model}
The discussion in the present paper is completely Newtonian. Stellar orbits
under consideration are bound to the central mass which determines their
form but the orbits are not exact ellipses for two reasons which
principally influence their long-term evolution, namely:

\begin{description}
\item[(i)] the gravitational field of the disc material acts on the star as
 a tiny perturbation;
\item[(ii)] successive interactions with an accretion disc also affect
 trajectories in an impulsive manner, when the star intersects the disc
 plane.
\end{description}
Each star is treated as a free test particle
moving in the combined gravitational field of the centre and the disc;
collisions with the disc material act as instantaneous periodic
perturbations of the trajectory. All other effects are beyond the scope
of the present paper (although they will have to be taken into account
in a future self-consistent model).

One can speculate that other effects, ignored in this paper, may have a
comparable result on stellar orbits. In particular, general relativistic
dragging of inertial frames due to rotation of the central object and
the disc material will break the spherical symmetry of the gravitational
field and result in sudden excursions of the mean orbital parameters. On
the other hand, gravitational radiation will result in a slow decay of
the orbit in a manner analogous to the effect of direct collisions of
the body of the star with the disc material. These effects also support our
conclusion that star-disc collisions should not be considered as the
only perturbation of stellar orbits when their long-term evolution is
discussed. Nevertheless, our very restricted choice to the two effects
listed above, (i) and (ii),
has an additional reason. Indeed, these effects are linked
to each other, because both of them are determined by the mass of the
disc: increasing the mass of the disc affects the stellar trajectories
more by its gravitational attraction, and, at the same time,
star-disc collisions also become more important (being on average
proportional to the surface density of the disc). We will demonstrate
that a consistent model involving any one of the two effects
{\em{}must\/} take the other effect into account too. But the main
novelty of this paper is in even a stronger claim: it is the first of
the effects mentioned above --- the gravity of the disc --- which influences
long-term evolution of the stellar orbits dominantly, while collisions
with the disc material represent an underlying mechanism causing
a slow and continuous orbital decay. In other words, changes
in eccentricity and inclination are driven dominantly by the disc
gravity and they can occur rather abruptly. From this perspective,
Rauch's (1995) statistical model of the star-cluster evolution due to
interactions with the central accretion disc 
requires also the inclusion of the influence of the disc gravity.

It is also worth mentioning that, in analogy with Rauch (1995), we
disregard, at this stage of the model, mutual interactions of the stars,
considering the star cluster as a collisionless system. A rigorous
approach will require us to solve the Fokker-Planck equation in a manner
analogous to Bahcall \& Wolf (1976, 1977; see also Peebles 1972; Young
1980; Shapiro \& Teukolsky 1985, 1986; Zamir 1993; Quinlan, Hernquist \&
Sigurdsson 1995; Sigurdsson, Hernquist \& Quinlan 1995.) Obviously, this
neglect represents a large simplification, especially in the nuclear
region close to the central galactic object (Statler, Ostriker \& Cohn
1987; Lee \& Ostriker 1993). Nevertheless, demonstration of
the effect of the disc
gravity upon stellar orbits which we want to discuss hereafter does not
call either for general theory of relativity nor mutual interactions
among stars themselves to be taken into account. We expect, however,
that the capture rate of stars by the central object can be only roughly
estimated in this approach (the capture rate has been discussed in
various approximation by, e.g., Frank \& Rees 1976; Nolthenius \& Katz
1982; Novikov, Pethick \& Polnarev 1992; Hameury et al.\ 1994;
Sigurdsson \& Rees 1997).

\subsection{Gravitational field}
We describe the gravitational field as a superposition of the spherically
symmetric field of the central object (potential $V_\BH(r)=-GM_\BH/r$;
$r$ is radial distance from the centre) and an axially symmetric field
of the disc (potential $V_\di({R},z)$; cylindrical coordinates
$({R},z)$, $z=0$ is the disc plane). The disc is geometrically thin and
it is described by surface density $\akpa(R)$. We assumed $\mu=10^{-3}$
for definiteness. It is worth mentioning that our $\akpa(R)$ corresponds
to vertically integrated density which is introduced in the standard
theory of geometrically thin discs. With this correspondence one can
compare our results with other works which employ Shakura--Syunaev
(1973) and Novikov--Thorne (1973) discs. We do not consider a more
complicated case of geometrically thick tori in this paper.

An analytical expression for the disc potential can be found for some
particular forms of the density distribution (Binney \& Tremaine 1987,
Evans \& de Zeeuw 1992). Despite the fact that such models can approximate
real astrophysical discs only roughly, their main advantage is the
analytical expressions for the gravitational field which they offer.
Obviously, a careful check is needed in order to verify that none of
the important qualitative features of the solution has been altered.
We shall follow this line of reasoning by working mainly with
highly simplified Kuzmin's class of discs.

The surface density--potential pair for Kuzmin's model reads
\begin{eqnarray}
 \akpa(R) &=& {M_\di A\over 2\pi}{1\over (A^2+R^2)^{3/2}}\; ,
  \label{ku1} \\
 V_\di({R},z) &=& -{G M_\di\over \left[{R}^2+\left(A+|z|\right)^2
  \right]^{1/2}}\; , \label{ku2}
\end{eqnarray}
where $A$ is a free parameter of the model. An easy exercise then
yields components of acceleration due to the disc gravitational field.
Kuzmin's discs are of infinite radial extent but their mass is finite
because the surface density decreases with radius fast enough.
Relevant formulae for discs of finite radial size and arbitrary
surface-density profile are summarized in Appendix.

In our numerical code for integration of stellar orbits we have
used either the simple Kuzmin formulae given above or, in case
of discs with a-priori unconstrained surface density distribution,
pre-computed $V_\di$ and components of the gradient
$(\partial{V_\di}/\partial{R},\partial{}V_\di/\partial{z})$ for a given
distribution of $\akpa$ in a fixed grid of $({R},z)$. Then we
employed six-point interpolation formulae and evaluated the gravitational
force acting on a star at any position. By performing several tests we
have tuned parameters of the grid in order to optimize computer time
necessary for integration of the stellar orbits and, simultaneously,
to preserve a pre-determined accuracy.

\subsection{Interaction with the disc material}
\label{interaction}
The physics of collisions of the stars with the disc material can be
represented by a prescription for the change of star's velocity. The
prescription is based on a simplified hydrodynamic scenario in which
the star crossing the disc is treated as a body in hypersonic motion
through fluid. The resulting change of velocity (which occurs always at
$z=0$, once or twice per each revolution) is obtained by integration over a
short period when star moves inside the disc material (as is done
with all quantities in the thin-disc approximation). Since the
pioneering works of Hoyle \& Lyttleton (1939), Chandrasekhar (1942), and
Bondi \& Hoyle (1944), the drag on a cosmic object has been considered
for various astrophysical problems. It has
been recognized that in the case of a supersonic flow which resembles our
situation (Ostriker 1983; Zurek, Siemiginowska \& Colgate 1994, 1996)
the hydrodynamic drag consists of a component given directly by the flow
of material onto the stellar surface (or into a stellar-mass black hole;
Petrich et al.\ 1989) and a long-range component
given by the interaction of the flow with a conical or a bow shock
surrounding the star. The relative importance of the two components is
sensitive to the details of the flow as well as to complicated turbulent
processes in the wake (e.g., Livio, Soker, Matsuda \& Anzer 1991; Zurek
et al.\ 1996). In what follows we will adopt a somewhat
simplified empirical model which, however, we argue
still reflects, rather conservatively, physical effects in our
consideration.

We adopt a simplified formula for the mutual interaction between star
and disc as in Vokrouhlick\'y \& Karas (1998). Conclusions drawn
from the present paper can be thus compared with previous results 
in which the gravity of the disc was ignored (see also Syer et al.\ 1991;
Artymowicz, Lin \& Wampler 1993; Rauch 1995). The impulsive change of 
the star's
velocity, $\delta\bmath{v}$, when it crosses the disc will be described
by

\begin{figure*}
 \epsfxsize=\hsize
 \centering
 \mbox{\epsfbox{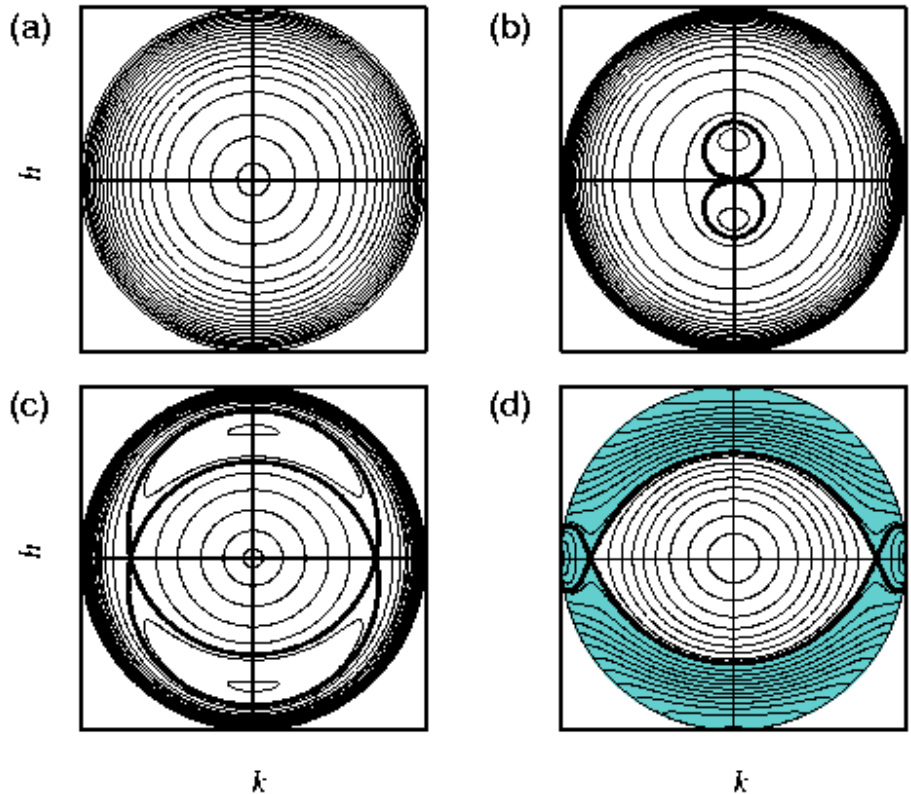}}
 \caption{Contours $C=\const$ of the first integral
  (\protect\ref{31two}) in the plane of non-singular variables $(k,h)$.
  Radial distance from origin determines mean eccentricity $\bar{e}$
  according to eqs.\ (\protect\ref{kh1})--(\protect\ref{kh2}). Polar
  angle (measured from the horizontal $k$-axis) is the argument of
  pericentre $\bar{\omega}$. A separatrix (thick curve) is drawn on the
  border between regions of different topology in panels (b)--(d); there
  is no separatrix in (a).
  Contour lines with polar angle acquiring values in the full range of
  $0\leq\bar{\omega}\leq2\pi$ correspond to a {\em{}zone of
  circulation}, while some values of $\bar{\omega}$ are not allowed in
  {\em{}zones of libration}. In this example, Kuzmin's disc with $A=75$
  (in units normalized by half of gravitational radius of the central mass)
  has been assumed. Graphs are shown for four values
  of Kozai's integral respresenting topologically different situations:
  (a)~$c=0.9$ ($0\leq\bar{e}\leq0.436$); (b)~$c=0.81$
  ($0\leq\bar{e}\leq0.586$) --- notice a figure of
  8-shaped separatrix between
  the libration and circulation regions; (c)~$c=0.7$ ($0\leq \bar{e}\leq
  0.714$) --- notice the appearance of the second (inner) circulation region
  around the origin; (d)~$c=0$ ($0\leq\bar{e}\leq1$). In all these cases,
  trajectories have semimajor axis ${\bar{a}}=200$. Shaded area
  indicates highly eccentric orbits which must be trapped by the central
  object at some stage. More details are given in the text. \label{f1}}
  \end{figure*}

\begin{figure*}
 \epsfxsize=\hsize
 \centering
 \mbox{\epsfbox{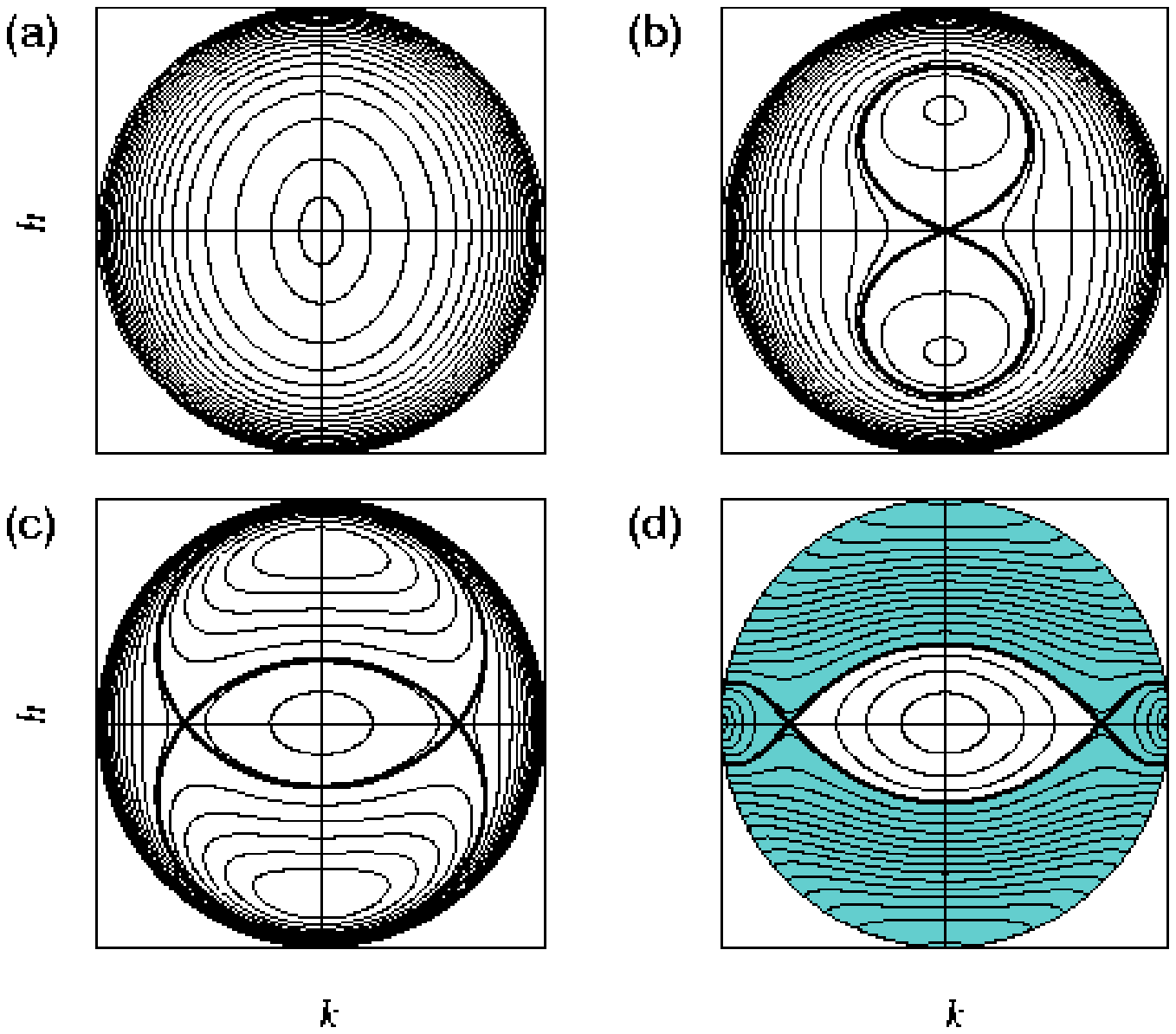}}
 \caption{The same as in Figure~\protect\ref{f1} but for orbits with
  a smaller semimajor axis, ${\bar{a}}=75$. Values of Kozai's integral are:
  (a)~$c=0.9$; (b)~$c=0.84$; (c)~$c=0.7$; (d)~$c=0$.
 \label{f2}}
\end{figure*}

\beq
 \delta\bmath{v}=\Sigma(R,\bmath{v})\,\bmath{v}_{\rm rel}
  \label{22one}
\eeq
with $\Sigma(R,\bmath{v})=
-\pi\zeta\akpa(R_\star^2/m_\star)(v_{\rm{rel}}/v_\perp)$, and
$\zeta\approx1+(v_\star/v)^4\ln\Lambda$. Here, $\bmath{v}_{\rm rel}$
denotes the relative velocity of the star with respect to the disc matter,
$R$ is the radial coordinate in the disc, $R_\star$ stellar radius,
$m_\star$ its mass and $v_\star$ the escape velocity from surface of the
star; $v_\perp$ is the normal component of the star's velocity to the disc
plane, and $\ln\Lambda$ is the long-range interaction factor. Obviously,
the latter term is to be considered for transonic flows only, i.e.\ when
the Mach number ${\cal M}\approx(R/H)>1$ (this condition is
well-satisfied in standard thin accretion discs of Shakura \&
Sunyaev 1973; cf.\ also Zurek et al.\ 1994 who
estimate the amount of the disc material swept out of the disc by a star's
passages). Finally, $H$ is the geometrical thickness of the accretion disc
at distance $R$. Ostriker (1983) gives a rough estimate of
$\Lambda\approx(H/R_\star)(v/v_\star)^2$. Again, for standard thin discs
one obtains $\Lambda\gg1$. We note again that the main concern is not to
underestimate the role of the hydrodynamic drag. We shall thus rather
conservatively assume that the factor $\zeta$ is equal to $10^3$.

It is worth mentioning that Artymowicz (1994) proved formula
(\ref{22one}) approximates also the interaction of the star with density and
bending waves excited on the disc surface by the star itself (see also
Hall, Clarke \& Pringle 1996). This additional effect can be
accommodated by an appropriate modification of the drag factor $\zeta$.
Taking into account an upper estimate on $\zeta$, we effectively include
this effect in our considerations too.

To conclude this paragraph, we recall that the factor $\Sigma$ in
eq.~(\ref{22one}) is proportional to the surface density and therefore
depends linearly on the total mass of the disc (given the density
profile). When expressed in units of the central mass:
$\Sigma\propto\mu$ with a numerical factor depending uniquely on the
characteristics of the moving object (neutron star, white dwarf,
stripped star, etc.) and of the disc material. Though simplified, this
model reflects the intuitive guess that more mass in the disc makes the
effects of the hydrodynamic drag more profound.

\section{Evolution of stellar orbits}
\label{evolution}
Hereinafter, we examine stellar orbits. First, we discuss how the
gravity of the disc influences individual trajectories (Sec.
\ref{gravity}). This will help us to illustrate the main differences
with respect to previous works (especially, Rauch 1995). Next, we
consider the combined effect of the disc gravity and the drag due to
collision with the disc material (Sec. \ref{drag}). In both cases, we
will integrate orbits numerically and then we will describe orbital
evolution in terms of osculating Keplerian elements. The elements
relevant for our work are: semimajor axis $a$, eccentricity $e$,
inclination $I$ with respect to the disc plane ($I>90^{\rm{o}}$
corresponds to retrograde orbits while $I<90^{\rm{o}}$ corresponds to
prograde orbits), and argument of pericentre $\omega$ as measured from
the ascending node. The longitude of the node does not appear in the
following discussion because of the axial symmetry of gravitational
field. The (assumed) small value of the disc-mass parameter $\mu$ and
the value of $\zeta$ guarantee that the time-scale of the orbital
evolution is much longer than period of a single revolution. Therefore,
we will average relevant quantities over individual revolutions around
the centre whenever it is appropriate. One can verify that $\mu$
controls the ratio of medium vs.\ short time-scale while $\zeta$ affects
the ratio of long vs.\ medium time-scale. We accept physically
substantiated values for $\mu$ ($\lta10^{-3}$) and $\zeta$ (1--$10^3$)
which guarantee that the three time-scales are well-separated from each
other.

\subsection{Effects of the disc gravity}
\label{gravity}
The motion of cosmic objects in gravitational fields of discs or rings
of matter has been considered in the context of solar system studies
(e.g., Ward 1981; Heisler \& Tremaine 1986; Lemaitre \& Dubru 1991;
McKinnon \& Leith 1995), recently discovered planetary systems (Holman,
Touma \& Tremaine 1997), and in galactic dynamics (Huang \& Carlberg
1997).

\begin{figure}
 \epsfxsize=1.108\hsize
 \mbox{\hspace*{-3ex}\epsfbox{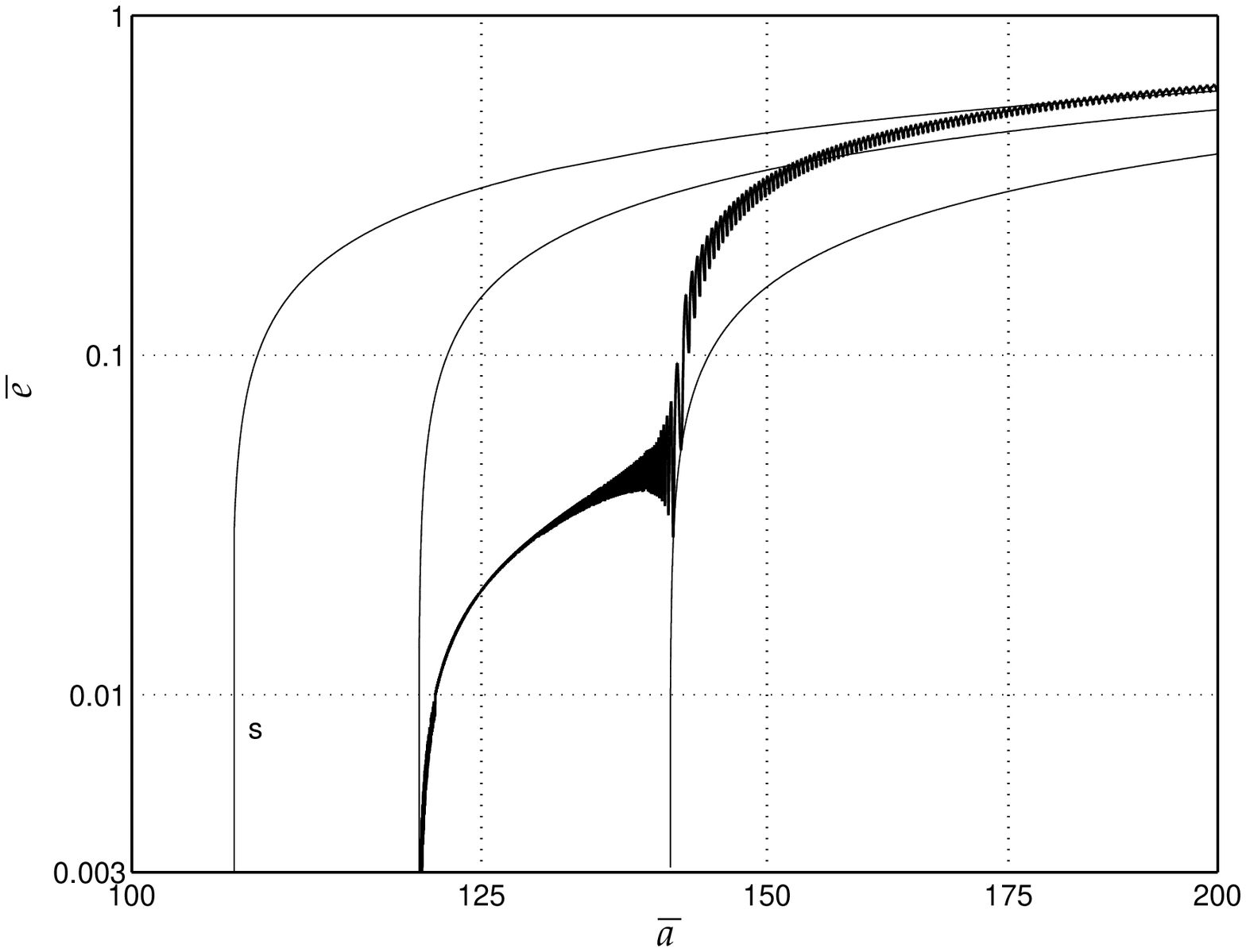}}
 \caption{Mean eccentricity $\bar{e}$ vs.\ semimajor axis $\bar{a}$.
  The curly curve corresponds to the trajectory from the first example in
  sec.~\protect\ref{drag}. Both axes are logarithmic. Kuzmin's disc
  model with scaling parameter $A=75$ and mass parameter $\mu=10^{-3}$
  has been used. Initial semimajor axis $\bar{a}=200$, eccentricity
  $\bar{e}=0.6$, inclination $\bar{I}=34.4\degr\!\!$, and argument of
  pericentre $\bar{\omega}=90\fdg$ For sake of comparison, three
  hypothetical orbits, evolution of which disregards gravitational
  influence of the disc, are also plotted (monotonic curves); one of these 
  curves (labeled by ``s'') corresponds to the same initial conditions as the
  complete solution.
 \label{f3}}
\end{figure}

As mentioned above, we are interested in the long-term variations of
orbits around a massive centre and a much less massive disc. The averaging
technique is a very useful approach for understanding the qualitative
behaviour of the system (Arnold 1989). It can be formalized in terms of
a series of successive canonical transformations in which rapidly changing
variables (e.g. mean anomaly along the osculating ellipse) are eliminated
(Brouwer \& Clemence 1961). Since the reader may not be closely
familiar with this technique we briefly remind several basic concepts.

Suppose we consider motion of a particle in the Keplerian field,
determined by an integrable Hamiltonian $H_0$ and a weak perturbation.
Perturbation is described by a potential $\epsilon V$. The parameter
$\epsilon$ indicates a smallness of the perturbation (in our case, the
disc/hole mass ratio $\mu$ plays the role of $\epsilon$). Hamiltonian of
the central field in terms of the Keplerian elements reads $H_0 =
-GM_\BH/(2a)$, depending uniquely on the semimajor axis, while the
perturbing potential $V$ depends, in general, on all Keplerian
parameters. For this reason, a general solution cannot be found but the
the problem is simplified when some symmetry appears. For instance, the
axial symmetry induces independence on the longitude of node. Since the
averaging technique is based on canonical transformation theory one
should rather use a set of canonical elements of the two-body problem.
Delaunay parameters are the most common choice (see, e.g., Brouwer \&
Clemence 1961).

In order to grasp the long-term evolution of the system one needs to get
rid of the ``fast'' (i.e.\ rapidly changing) variables. In our problem
there is the only one fast variable: the mean anomaly. The idea of the
averaging method is then formalized by seeking a set of new Delaunay
variables using a canonical transformation under the constrain of
independence of the transformed Hamiltonian on this fast variable. It is
possible, indeed, by formal development in the small parameter
$\epsilon$. The original perturbing potential $V$ reads, after the
transformation,

\begin{equation}
 \epsilon V\; \rightarrow\; \epsilon {\bar V} = \epsilon {\bar V}_1 +
  \epsilon^2 {\bar V}_2 + \epsilon^3 {\bar V}_3 + \cdots \; .
 \label{ave1}
\end{equation}
Various iterative and computer-algebra adapted methods for recursive
generation of the potentials ${\bar V}_1$, ${\bar V}_2$, etc.\ were
developed (e.g.\ Hori 1966; Deprit 1969). We note that the first
term in this series, notably ${\bar V}_1$, is just the
{\em{}average of the original potential $V$ over the mean anomaly}.
Obviously, a rigorous proof of convergence of the series (\ref{ave1}),
and thus a success of the whole procedure, remains a very difficult
task. Despite of this fact the averaging technique is often very useful.
Typically, results based even on highly truncated part of this series
are valid on a limited time interval though they fail to predict the
system's evolution on an infinite time-scale. In many applications such
a weakened demand is sufficient (see, for instance, Milani \&
Kne\v{z}evi\'c 1991 for the application to a long-term evolution of
asteroidal orbits).

Regarding the independence of the new Hamiltonian on the fast variable
(i.e.\ mean anomaly) we immediately conclude that the mean
perturbing potential ${\bar V}$ is the first integral of motion:

\beq
 \epsilon {\bar V} = \epsilon {\bar V}_1 + \epsilon^2 {\bar V}_2 +
  \epsilon^3 {\bar V}_3 + \ldots = \epsilon C(\epsilon) \; .
 \label{ave2}
\eeq
Factorizing out the small parameter $\epsilon$ one can see that the
first order perturbing function, ${\bar V}_1$, is nearly constant ---
provided that the higher order contribution is neglected. Notice that
the right-hand side constant $C$ is a function of the small parameter
$\epsilon$. However, in the most truncated level of averaging,
accounting for the first order term of the right hand side of
(\ref{ave1}) only, we have

\beq
 {\bar V}_1 \approx C(0) + {\cal O}(\epsilon)\; , \label{ave3}
\eeq
and the averaged perturbation potential ${\bar V}_1$ is (approximately)
constant. In the following we shall drop out the argument (which is
zero) and write $C$ for simplicity. In the case of an axially symmetric
system the integral (\ref{ave3}) is sufficient for global integrability.
Now the problem is reduced to a single degree of freedom and the whole
phase space can be plotted in a simple two-dimensional graph.
Qualitative features of the solution can be illustrated by isocurves of
the $C$-integral (\ref{ave3}).

In the remaining part of this section we shall apply the previous
brief review of the averaging to our problem. We shall restrict
ourselves to the first-order procedure in which the perturbation
potential $V_\di$ is substituted by its average $\bar{V}_\di$ over 
the mean anomaly:
\beq
\bar{V}_\di=\frac{1}{2\pi\eta}\int_0^{2\pi}{\rm{d}}v
\left(\frac{r}{\bar{a}}\right)^2 V_{\rm{d}}({R},z),
\eeq
where $r$, ${R}$, and $z$ are functions of true anomaly $v$,
$\eta=\sqrt{1-\bar{e}^2}$ ($\bar{e}$ denotes mean eccentricity of
the orbit). One can write, in terms of mean
inclination $\bar{I}$ and mean argument of pericentre
$\bar{\omega}$,

\begin{eqnarray}
r &=& \frac{\bar{a}\eta^2}{1+\bar{e}\cos v}, \\
z &=& r\sin\bar{I}\sin\left(\bar{\omega}+v\right), \\
{R} &=&
r\sqrt{1-\sin^2\bar{I}\sin^2\left(\bar{\omega}+v\right)}.
\end{eqnarray}
At this level of approximation, the
mean semimajor axis $\bar{a}$ stays constant and
differs from the osculating semimajor axis $a$ by short-period terms
only. Owing to axial symmetry of $\bar{V}_\di$ there exists an
additional first integral of motion which relates
${\bar{I}}$ to the corresponding value of
${\bar{e}}$ in the course of their evolution,

\beq
 \sqrt{1-\bar{e}^2}\cos\bar{I}=c\equiv\const\,.\label{31one}
\eeq
Eq. (\ref{31one}) is often called Kozai's integral
(Kozai 1962). Here again, the overbar
distinguishes the mean elements from the corresponding osculating
elements. The problem
is reduced to the evolution of ${\bar{e}}$ and
${\bar{\omega}}$ which are constrained by

\beq
 {\bar V}_\di\left({\bar{e}},{\bar{\omega}};c,{\bar{a}}\right)=C\,.
 \label{31two}
\eeq
We will introduce a pair of non-singular canonical variables $(k,h)$:

\begin{eqnarray}
 k &=& \sqrt{2(1-\sqrt{1-{\bar{e}}^2})}\cos{\bar{\omega}},
 \label{kh1} \\
 h &=& \sqrt{2(1-\sqrt{1-{\bar{e}}^2})}\sin{\bar{\omega}};
 \label{kh2}
\end{eqnarray}
$(k,h)$ are often called Poincar\'e variables, since
they have been introduced to orbital dynamics by Poincar\'e (1892).
Levels of $C={\const}$ in the $(k,h)$-plane offer a
convenient representation of the long-term evolution of mean
orbital elements. At small values of eccentricity ${\bar{e}}$,
the radial distance
from the origin in the $(k,h)$-plane is equal to the mean eccentricity
itself, while the polar angle has the meaning of the argument of pericentre
${\bar{\omega}}$. The final task is to evaluate the averaged
potential, ${\bar{V}}_\di\left({\bar{e}},{\bar{\omega}};c,{\bar{a}}
\right)$. Since we are interested in averaging over orbits which
intersect the disc, no simple analytical techniques based on expansion
in orbital elements (common in celestial mechanics) can be applied
to a general distribution of surface density $\akpa(R)$
(see the Appendix). Even in the case of simple Kuzmin's discs (\ref{ku2})
the averaging cannot be performed in analytical functions and
must be obtained numerically.

Figure~\ref{f1} illustrates contours $C={\const}$ of the first integral
(\ref{31two}) in the case of a Kuzmin disc with scaling parameter
$A=75$. The size of orbit is characterized by a value of mean semimajor
axis ${\bar{a}}=200$. Geometrized units of length have been used; in the
following, we will use half of the gravitational radius of the central
object as a natural unit of length (this choice is motivated by
interpreting the central object as a black hole; otherwise, our
discussion is Newtonian).

Kozai's integral $c$ spans the whole interval $(-1,1)$, but we can restrict
ourselves to
positive values because $c$ occurs in ${\bar V}_\di$ only squared.
Large values of $c$ (Fig.~\ref{f1}a, $c=0.9$) correspond to
quasi-circular orbits with low inclination to the disc plane and small
oscillations in mean eccentricity. The argument of pericentre circulates
in the whole interval $(0,2\pi)$. When $c$ is decreased (Fig.~\ref{f1}b,
$c=0.7$), larger inclinations occur and two regions of libration with
associated stable points at $\bar{\omega}=\pm\pi/2$ develop and
bifurcate into a figure of 8-shaped region. Orbits outside this libration region
still circulate with $0\leq\bar{\omega}\leq2\pi$ ($0$ is to be identified
with $2\pi$). Notice that circulating trajectories which are close to
the separatrix of the two regions exhibit large oscillations in the mean
eccentricity. At still smaller values of $c$ (Fig.~\ref{f1}c, $c=0.2$),
the circulation region bifurcates in its inner part near origin. Stable
points in the libration regions are expelled farther to higher
eccentricities. Setting $c=0$ (Fig.~\ref{f1}d) the inclination is
constrained to $\bar{I}=\pi/2$ (polar orbits). There is a maximum
eccentricity above which the orbits are trapped by the central object
(this situation corresponds to a pericentre distance equal to $2$ in our
example). We observe that a large portion of the plot (indicated by
shading) corresponds to orbits which emerge from or fall into
the centre, while only a very minor part --- namely the inner
circulation region with a stable point $\bar{e}=0$ --- contains orbits
which survive the long-term evolution. In the following, this property turns
out to be essential for the fate of orbits counter-rotating with respect
to the disc. We will argue that most initially retrograde orbits
cannot last long enough to be tilted over the polar orbit and inclined
into the disc plane because, in course of this process, their
eccentricity is being pumped up to such a high value that they get
captured by the central object.

\begin{figure*}
 \epsfxsize=\hsize
 \centering
 \mbox{\epsfbox{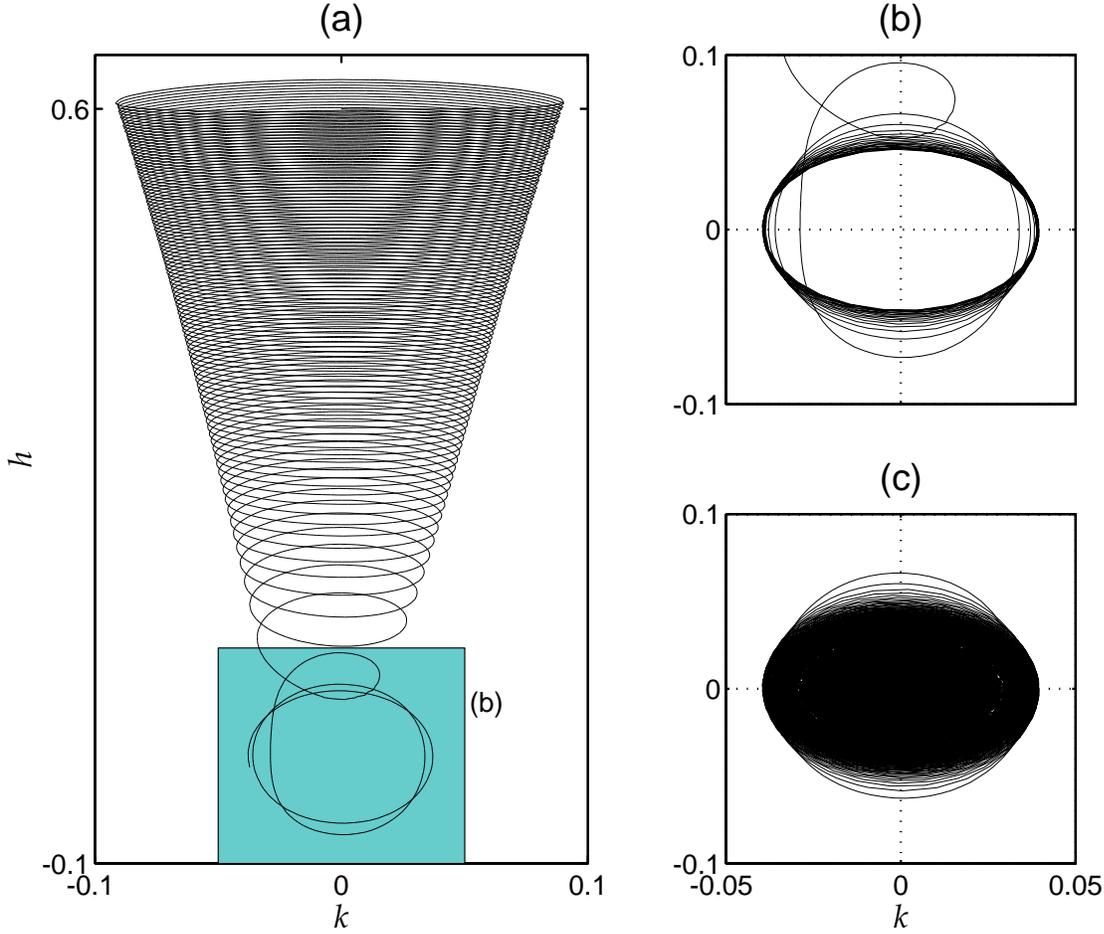}}
 \caption{Long-term evolution of the orbit from Figure~\protect\ref{f3}
  projected onto the $(k,h)$-plane. Initial librations around the fixed
  point at $\bar{\omega}=90\degr$ are shown in panel (a). As the Kozai's
  parameter $c$
  increases, the libration zone shrinks and the orbit approaches the
  separatrix. Eventually, the orbit crosses the separatrix and starts
  circulating around the origin; see continuation of orbital evolution
  in panel (b). Still further continuation in (c) shows that the orbit
  terminates in the inner zone of circulation, with zero eccentricity.
 \label{f4}}
\end{figure*}

Figure~\ref{f2} shows the same graphs as in Figure~\ref{f1} but for
different ratio of the scaling parameters of density distribution
($A=75$, as before) and the orbit ($\bar{a}=75$, changed). All features
in the $(k,h)$-plane described before stay unchanged except the fact
that libration regions and the inner zone of circulation cover a larger
portion of the graphs. At small values of Kozai's integral we again
observe that eccentricity is forced to large oscillations which
eventually lead to capture (shaded area).

We have verified that qualitatively similar results hold for
different, astrophysically relevant disc density--potential
pairs. In particular, using formulae from the Appendix we have
numerically computed potential of discs with $R^{-1}$, $R^{-2}$ and
$R^{-3}$ surface density laws and then averaged over corresponding
quasi-elliptic orbits.

\subsection{Influence of star-disc collisions}
\label{drag}
We now consider inclusion in our model of the influence of the
hydrodynamic drag affecting the orbit when the star crosses the
accretion disc. We will use the approximations described in
Sec.~\ref{interaction}, and, for definiteness, we will assume that the
central cluster of stars is formed by white dwarfs or stripped cores of
ordinary stars and use relevant parameters for estimating numerical
factor in the relation $\Sigma\propto\mu$ (eq.~[\ref{22one}] above).
Before embarking on a description of our results, we recall that Rauch
(1995) has considered long-term evolution of stellar orbits in which he took
into account the impulsive drag acting twice per revolution only
(no gravity of the disc; see also previous studies by Syer et al.\
1991, and Vokrouhlick\'y \& Karas 1993, 1998).

All stellar orbits interacting with the disc, independently of their
initial conditions, exhibit long-term decay of the semimajor axis, and
monotonous decrease of eccentricity and inclination (grinding to the
disc plane). For initially low-inclination orbits, the characteristic
time-scale of circularization is comparable to the grinding time after
which the orbital plane becomes inclined into the disc. Orbits with
large initial inclination, in particular all initially retrograde
orbits, are circularized before they incline to the disc. Rauch (1995)
argues that, for orbits with an initially moderate eccentricity, a
particular combination of the mean elements,
$\bar{a}(1-{\bar{e}}^2)\cos^4({\bar{I}}/2)$, is quasi-conserved in
the course of evolution (see also Vokrouhlick\'y \& Karas 1998). This
property is insensitive to a particular model of the star-disc
interaction, surface density profile, and the total mass of the disc.
For further use we recall the way that the hydrodynamical drag
affects the mean semimajor axis ${\bar{a}}$ and the Kozai parameter $c$
(two integrals of the long-term evolution when star-disc interactions are
neglected): due to the drag effect, $\bar{a}$ undergoes permanent
decay while $c$ typically increases from its initial value to the final
value of $c_{\rm f}=1$, corresponding to a circular orbit in the disc
plane. Initially retrograde orbits behave in a somewhat different way:
their circularization time-scale is significantly shorter than the grinding
time. For these orbits, the value of Kozai's parameter first slightly
decreases before exhibiting monotonous increase to $c_{\rm{}f}$.

{\em{}The main result of the present paper is that most of the
above-mentioned properties cease to be true in our generalized model
which, besides star-disc collisions, includes also the gravitational
influence of the disc matter}. The key point is the fact that the
characteristic time-scale
$\bar{a}/\dot{\bar{a}}\approx\bar{c}/\dot{\bar{c}}$ of the
hydrodynamical-drag effects is much {\em{}longer\/} than the
characteristic time-scale for circulation and libration of pericentre.
Circulation and libration along $C=\const$ lines are due to the disc
gravity, and the corresponding period is also the time-scale for
gravity-driven oscillations in eccentricity and inclination. As a
consequence, the principal features of orbital variations explained in
the previous section remain unchanged apart from a very slow adiabatic
evolution of quasi-integrals $(\bar{a},c)$. However slow this evolution
is, stellar orbits are strongly and abruptly affected at some stages:
when they cross the separatrix (due to some perturbation) or when the
libration region bifurcates. In the following paragraphs we will
illustrate these facts by showing typical orbits. Kuzmin's disc with
$A=75$ has been assumed in examples described below. The mass of the
disc has been set to $\mu=10^{-3}$ for definiteness (units of the
central mass). It should be mentioned that the results depend only very
weakly on the particular value of $\mu$, provided it is sufficiently
small. The reason emerges from the fact that the disc-mass parameter can
be factorized out of all expressions containing the averaged potential
${\bar V}_{\rm d}$ (which determines all important dynamical features).
On the other hand, and in contrast to the results of Rauch (1995), one
expects dependence on the surface density profile $\akpa(R)$.

\begin{figure}
 \epsfxsize=1.108\hsize
 \mbox{\hspace*{-3ex}\epsfbox{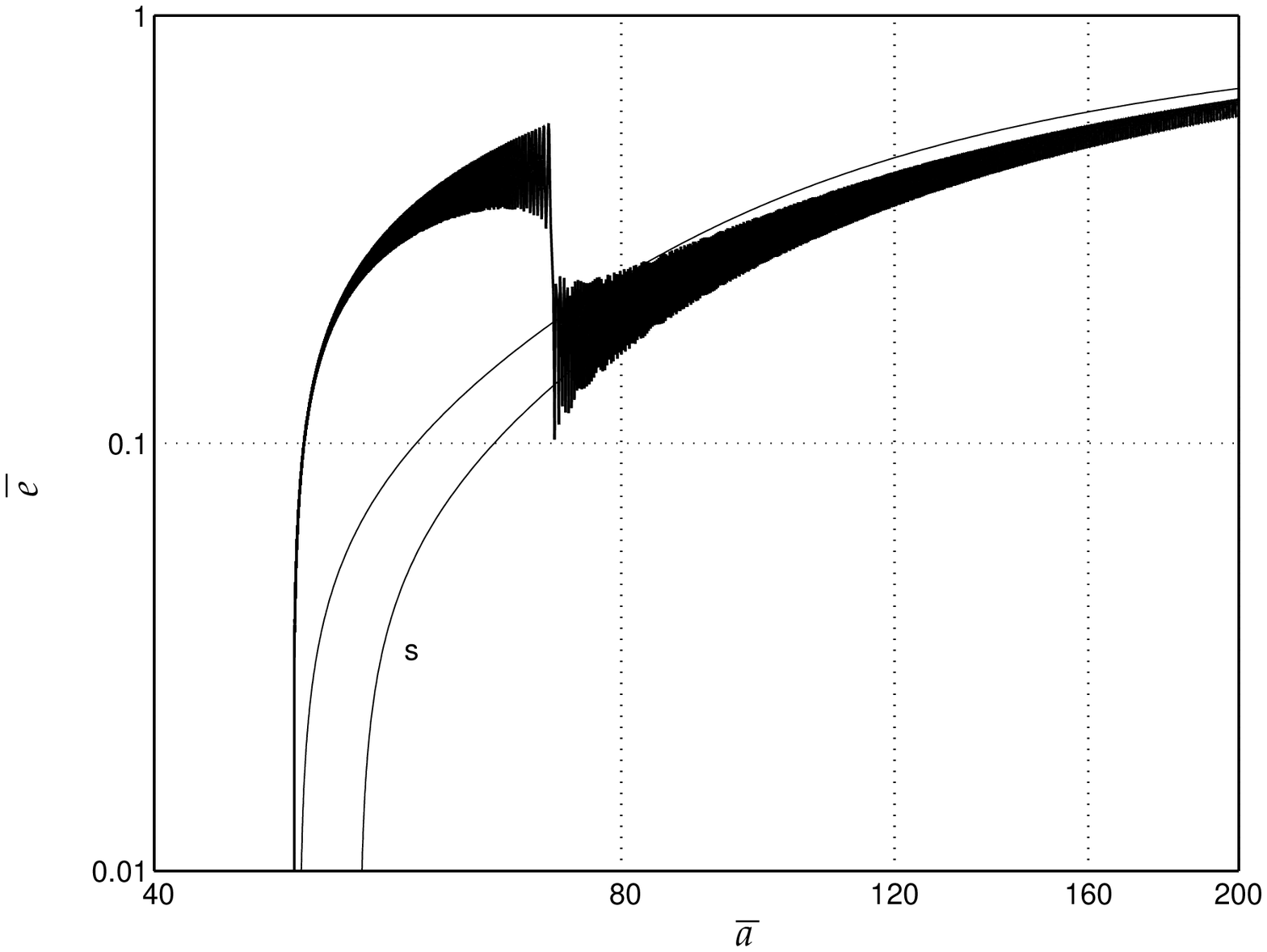}}
 \caption{Eccentricity $\bar{e}$ vs.\ semimajor axis $\bar{a}$
  (the second example in sec.~\protect\ref{drag}). The same parameters of the
  disc as in Figure~\protect\ref{f3}, different initial parameters of
  the orbit: $\bar{a}=200$, $\bar{e}=0.67$, $\bar{I}=67\degr\!\!$, and
  $\bar{\omega}=0$. Again, the gravitation of the disc induces oscillations
  and an abrupt increase of eccentricity (curly curve) which cannot be
  seen when the disc gravity is neglected (two monotonic curves).
 \label{f5}}
\end{figure}

In the first example of this section, we consider an orbit with the
initial parameters: $\bar{a}=200$ (in units of one half of the
gravitational radius), $\bar{e}=0.6$, $\bar{I}=34.4\degr$
($c\approx0.66$), and $\bar{\omega}=90\fdg$ Figure~\ref{f3} shows mean
eccentricity $\bar{e}$ as a function of the semimajor axis $\bar{a}$
during the course of orbital evolution. For sake of comparison we have
also plotted three curves of orbital evolution if the gravitational
influence of the disc is ignored. These curves can be compared directly
with corresponding results of Vokrouhlick\'y \& Karas (1993) and Rauch
(1995). The curve labeled by ``s'' corresponds to the same initial
conditions as the orbit of the complete model with gravitational effects
taken into account (notice the difference in predicted radius of the
terminal orbit); the other two curves correspond to different initial
eccentricities. As expected from the previous discussion, the most
evident features are (i)~large oscillations in eccentricity, and
(ii)~permanent decay of the semimajor axis. The radius of the resulting
circularized orbit in the disc plane, $a_{\rm{}f}=120.2$, differs
slightly from the estimation based on Rauch's (1995) formula,
$a_{\rm{}f}^{\rm{R}}\approx{}a_{\rm{}i}(1-e_{\rm{}i}^2)\cos^4{}I_{\rm{}i}/2=
106.5$. A closer look at orbital evolution explains a remarkable bump in
eccentricity near $\bar{a}\approx141$. In the initial state, the
pericentre of the orbit librates in one of the two lobes of the 8-shaped
zone, around $\bar{\omega}=90\degr\!\!$, as shown in Figure~\ref{f4}a.
As the Kozai parameter $c$ slowly increases due to the hydrodynamic drag
the libration zone shrinks and the orbit approaches the separatrix. At a
particular instant, the orbit is expelled from the libration zone,
crosses the separatrix and enters the zone of circulation (see
Figure~\ref{f4}b). This transition is accompanied by large oscillations
of the mean eccentricity and temporal slowing down of the eccentricity
decay.

\begin{figure*}
 \epsfxsize=\hsize
 \centering
 \mbox{\epsfbox{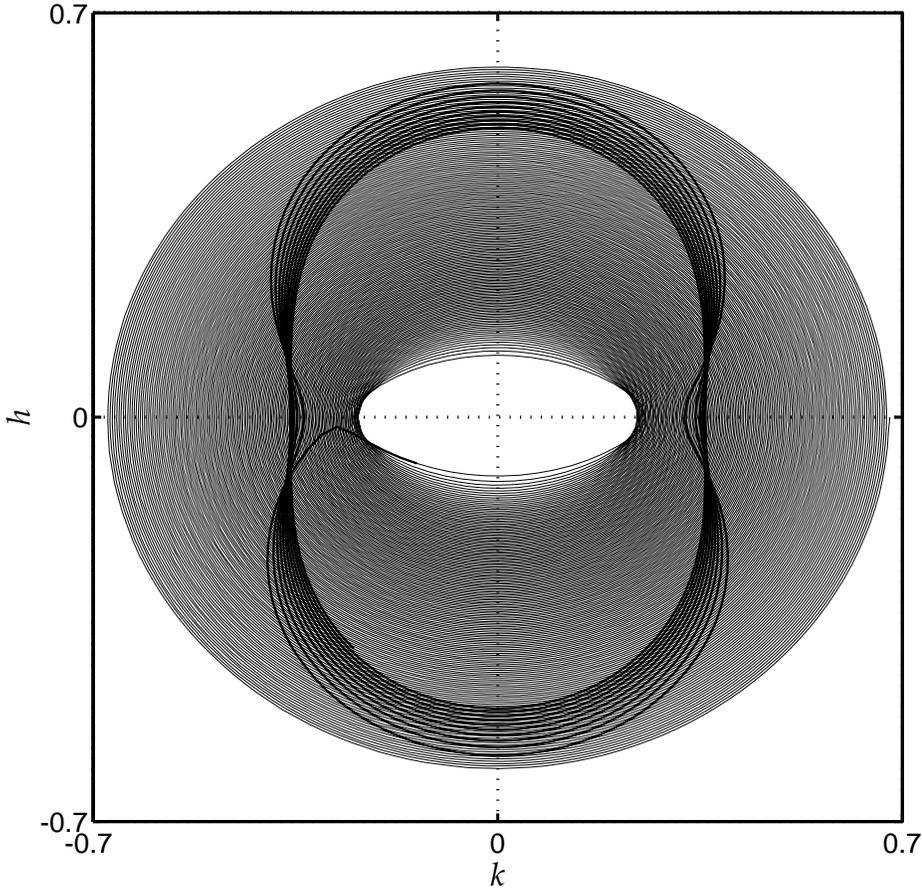}}
 \caption{Orbital evolution from Figure~\protect\ref{f5} projected onto the
  $(k,h)$-plane. Initial circulation in the inner zone of
  circulation (thin curve) is followed by transition to the external
  zone of circulation (thick curve) at a critical value $c\approx0.78$ of the
  Kozai parameter. At this moment eccentricity is increased back
  to a high value. Notice also a reversal of the sense of circulation.
 \label{f6}}
\end{figure*}

\begin{figure}
 \epsfxsize=1.108\hsize
 \mbox{\hspace*{-3ex}\epsfbox{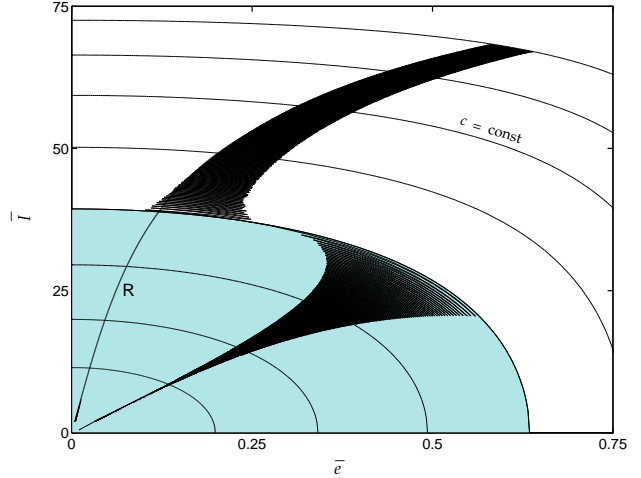}}
 \caption{Mean inclination $\bar{I}$ vs.\ mean eccentricity $\bar{e}$ of
  the orbit from Figure~\protect\ref{f5}. Background grid of constant
  values of the Kozai parameter, $c=\const$, is also plotted. Orbital
  evolution consists of fast oscillations along the lines of $c=\const$
  and slow diffusion towards the maximum value of $c_{\rm{}f}=1$
  ($\bar{e}=\bar{I}=0$). Kozai's quasi-integral is not isolating because
  the transition between the inner and the outer (shaded) circulation zones
  crosses one of the $c=\const$ curves. Rauch's solution is also shown
  by the monotonic curve (labeled by ``R'').
  \label{f7}}
 \end{figure}

Our second example of the orbital evolution (Figure~\ref{f5}) appears
even more peculiar (from the viewpoint of previous results neglecting
gravity of the disc). Initial orbital parameters are as follows:
$\bar{a}=200$ , $\bar{e}=0.638$, $\bar{I}=67\degr$ ($c\approx 0.3$), and
$\bar{\omega}=0$. The first part of the orbital evolution ends at
$a\approx 71.5$ by a significant increase of mean eccentricity. 
The oscillations of eccentricity eventually settle down and the evolution
terminates as a circular orbit in the disc plane, $a_{\rm{}f}
\approx{}49.2$. The latter radius is to be compared with Rauch's
approximative result: $a_{\rm{}f}^{\rm{R}}\approx{}54.3$. Obviously,
this estimate fails to predict the final radius correctly but the
difference is still within uncertainty of Rauch's quasi-integral
(Vokrouhlick\'y \& Karas 1998). Again, a closer look at the evolution of
pericentre in the $(k,h)$-plane sheds light on properties of this model.
Initially the orbit is confined to the inner zone of circulation around
the origin (Figure~\ref{f6}). An adiabatic increase of Kozai's
parameter results in collapse of this zone and the orbit is expelled
towards the separatrix. At the instant of crossing the separatrix, the
orbit may either enter the libration zone surviving for larger values of
$c$, or it terminates in the large circulation zone. It can be argued,
that orbits close to the separatrix spend most of their time near the
hyperbolic points which leads preferentially to the capture in the
circulation zone. This indeed happened in our example, as indicated by a
thick line in Figure~\ref{f6}. The form of the 8-shaped libration zone
results in significant increase of eccentricity. Figure~\ref{f7} shows
a projection of the orbit onto the plane of mean eccentricity $\bar{e}$
and inclination $\bar{I}$. One can notice that individual oscillations
are confined to the underlying grid of constant Kozai's integral
(\ref{31one}). Slow (adiabatic) diffusion across the lines $c=\const$
reflects a long-term feature of the orbital evolution. Transition from
the inner to the outer circulation zones is accompanied by a large
increase of eccentricity, but only a moderate decrease of inclination.

So far we have demonstrated the intricate role of the disc gravity in
evolution of stellar orbits with initially {\em{}prograde\/} inclination
($I_{\rm{}i}<90\degr$). Even though details of the orbital evolution are
different when compared to models neglecting disc gravity, overall
features remain approximately unchanged. Most importantly, radii of
circularized orbits in the disc plane are comparable. However, as
mentioned before, one has to pay particular attention to orbits with
initially large, {\em{}retrograde\/} inclination where the results
become qualitatively and quantitatively different. Two subsequent
examples deal with this class of orbits.

\begin{figure}
 \epsfxsize=1.108\hsize
 \mbox{\hspace*{-3ex}\epsfbox{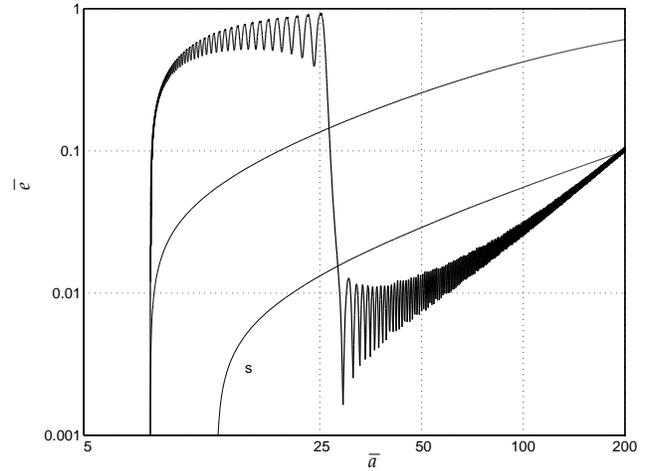}}
 \caption{Mean eccentricity $\bar{e}$ vs.\ semimajor axis $\bar{a}$ (the
  third example in the text; an initially retrograde orbit). The same
  parameters of the disc as in Figure~\protect\ref{f3}; starting
  parameters of the orbit are as follows: $\bar{a}=200$, $\bar{e}=0.1$,
  $\bar{I}=120\degr\!\!$, and $\bar{\omega}=90\fdg$ Complete model orbit
  is represented by a curly curve with oscillations in eccentricity,
  while the two orbits of a simplified model (no gravity of the disc)
  exhibit monotonic decrease of eccentricity during the whole
  evolution.
 \label{f8}}
\end{figure}

\begin{figure*}
 \epsfxsize=\hsize
 \centering
 \mbox{\epsfbox[017 380 593 611]{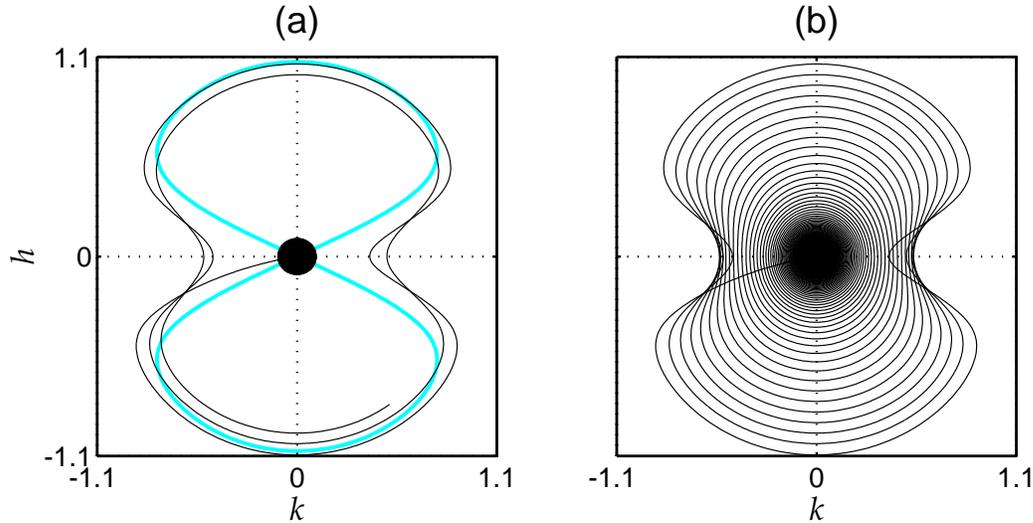}}
 \caption{Long-term evolution of the orbit from Figure~\protect\ref{f8}
  projected onto the $(k,h)$-plane. Evolution in the inner zone of
  circulation is shown in panel (a), until the trajectory escapes to the
  outer zone of circulation. Form of the 8-shaped separatrix is
  indicated at the moment of transition to high eccentricity.
  Subsequent evolution continues in panel (b) with a steady decrease of
  eccentricity. The trajectory eventually approaches the
  origin of the $(k,h)$ plane.
 \label{f9}}
\end{figure*}

Figure~\ref{f8} corresponds to an initially retrograde orbit with
parameters: $\bar{a}=200$, $\bar{e}=0.1$, $\bar{I}=120\degr$ ($c\approx
-0.5$), and $\bar{\omega}=90\fdg$ Similar to the previous example,
this orbit is originally locked in the inner zone of circulation. This
is necessary (but insufficient as we shall see later) if the orbit is to
be tilted over the polar orbit to a prograde one and survive further
evolution. Otherwise, it gets captured rather soon by the central mass.
Just before the separatrix of the inner circulation zone shrinks to
origin, the orbit is released to the outer zone of circulation.
The existence of the libration lobes then leads to significant increase of
the mean eccentricity, up to $0.9$. The $(k,h)$-plane representation of
the orbit evolution is given in Figure~\ref{f9}. The circular equatorial
orbit with radius $a_{\rm{}f}\approx{}7.8$ is a final state of
this evolution. Its radius is to be compared with
$a_{\rm{}f}^{\rm{R}}\approx{}12.4$ (no gravity of the disc). The
difference
between terminal radii predicted by the two models amounts to $30\,$\%.
Starting with various initial conditions we found that terminal radius
and corresponding time are comparable, typically within factor of 2.

\begin{figure}
 \epsfxsize=1.108\hsize
 \mbox{\hspace*{-3ex}\epsfbox{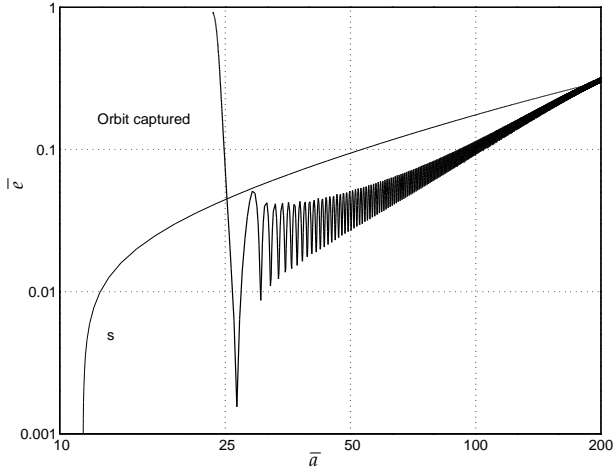}}
 \caption{Eccentricity $\bar{e}$ vs.\ semimajor axis $\bar{a}$
  (the fourth example in sec.~\protect\ref{drag}; initially retrograde
  orbit). The same disc and orbit parameters as in
  Figure~\protect\ref{f8} except for the initial mean eccentricity
  $\bar{e}=0.3$. The orbit which follows from our complete model has
  been captured by the central mass due to a large increase of
  eccentricity at the moment of transition from the inner to the outer
  zone of circulation, while the simplified model orbit grinded down
  safely to the equatorial plane.
 \label{f10}}
\end{figure}

A truly fundamental difference between the complete model, with the disc
gravity taken into account, and the simplified model, disregarding
effects of the disc gravity, is observed when the initial eccentricity
of the previous orbit is slightly increased to $0.3$. The corresponding
orbital evolution is shown in Figure~\ref{f10}. The
orbit is again initially locked in the inner zone of circulation.
But now, at the transition to the outer circulation zone,
eccentricity increases over a critical level and the orbit is captured
by the central object (pericentre is less than $2$). On the contrary, a
hypothetical orbit with no gravity of the disc grinds to the disc plane
at final radius $a_{\rm{}f}\approx{}11.4$. Concerning retrograde orbits,
we can conclude that even those ones that have started within the inner
zone of circulation are not safe from capture (obviously, orbits with
larger initial eccentricity than $0.3$ in our example are also
captured). We have also verified that retrograde orbits which are
initially locked in the libration lobes or in the outer circulation zone
do not survive tilting over the polar orbit, being soon captured by the
centre due to large oscillations in eccentricity. All
these orbits are grinded safely to the equatorial plane in the
framework of the simplified model of Rauch (1995), when the disc gravity
is neglected. We would like to stress again that this difference is not
based on the particular value of the disc mass we have chosen in our
examples ($\mu=10^{-3}$); rather it is present for an arbitrary nonzero
mass of the disc (naturally, evolution takes place on a longer
time-scale for smaller values of $\mu$). The same results hold also for
less massive discs.

A theoretical substantiation for the reported $\mu$-independence of our
results is based on existence of the quasi-integral (\ref{ave3})
mentioned above. Notice that the small parameter $\epsilon$ (i.e.\ $\mu$
in our case) can be factorized from this formula. Only when the
higher-order terms of the exact integral (\ref{ave2}) are taken into
account the results (phase portraits, in particular) depend on $\mu$. As
soon as $\mu$ is sufficiently small, which means smaller than about
$10^{-3}$ in practice, the averaging technique offers a simple
explanation for the $\mu$-insensitivity of the results.

We have verified our principal conclusions for discs with different
surface density profiles. In particular, we considered stellar orbits around
discs with density profiles proportional to $R^{-1}$, $R^{-2}$, and
$R^{-3}$. Obviously, the values of parameters when transitions between
zones in the $(k,h)$-plane occurred are quantitatively different in
individual cases, but what holds unchanged are the qualitative results.
Most importantly, a significant fraction of the initially retrograde
orbits is captured by the central mass.

\subsection{Notes on statistical properties of the cluster}
In this section, we briefly comment on possible influence of our results
on statistical properties of a cluster of stars near a galactic nucleus.

The important finding, in this respect, concerns the significant portion
of orbits which increase eccentricity and than become captured by
the centre, many of them being initially retrograde. These orbits are
missing in the final configuration of the system. The form of this final
configuration appears to be sensitive to the detailed nature of the
situation under consideration. Rauch (1995) specified the initial
distribution of stars as a power-law in binding energy. Then the system
evolved under interactions with a disc for infinite time, stellar orbits
were inclined into the disc plane or captured by the centre, and the
index of the power-law has been evolved accordingly. No new stars have
been inserted into the system. Since retrograde orbits end up with
smaller radii in the disc plane than initially prograde ones [due to the
term $\cos^4(I/2$) in Rauch's estimate], we expect in our model
deficiency of orbits very close to the central object when compared to
the results of Rauch. As a consequence a softer power law of the final
distribution is to be expected in comparison to the Rauch's work.

Alternatively, one can look for equilibirum in which the number density of
stars remains constant at all radii. Stars which get captured by the
central mass must be substituted by new ones which come from infinite
radius. Now it is important to take into account the fact that time
interval for grinding the orbit into the disc depends on initial
inclination (it is longer for initially retrograde orbits than for
prograde). Discussion in this section is to be verified by detailed
numerical simulations (work in progress).

\section{Conclusions}
\label{conclusions}
It has been recognized by previous works that statistical properties of
central galactic clusters are influenced by an accretion disc
surrounding the nucleus because of twice-per-revolution interaction
which affects stellar motion. The main results of this paper can be
summarized as follows:

\begin{description}
\item[(i)] we demonstrated that any consistent model
 of the star-disc interaction has to take the influence of the disc gravity
 into account, in addition to the effects of direct collisions with gaseous
 material;
\item[(ii)] as a result of the disc gravity, individual stellar orbits
 exhibit evolution which is  different
 if compared to the situation when collisions are considered
 but gravity neglected. Most importantly, we found that a significant fraction
 of initially retrograde (i.e.\ counter-rotating with respect to the
 disc material) orbits are captured by the central object. This is due
 to large oscillations in eccentricity which affect polar orbits.
\end{description}

We wish to note that our two findings mentioned above are to some extent
different in their nature. The former one --- (i) --- is essentially a
statement of consistency claiming that any reasonable model which
involves the effects of direct star-disc physical interaction has to
take disc gravity also into account. We argued that this claim is valid
for all astrophysically reasonable objects expected in central clusters
of galaxies: neutron stars, white dwarfs and stripped stars. The logic
behind this result is due to the fact that both effects are controlled
by the total mass of the disc. The latter finding --- (ii) --- then
states how the model supplemented by effects of the disc gravity differs
from previous simpler models. Gravity of the disc induces dynamical
structures, libration and circulation zones of the argument of pericentre.
Adiabatic change of quasi-integral quantities and related transitions of
trajectories between the two zones are the essence of our results.

It is worth recalling that the above-mentioned results did show sensitivity
on a particular model of the disc, especially on the radial gradient of the
surface density. Indeed, while Rauch (1995), considering only star-disc
collisions, reported his results to be insensitive to a particular
profile of the surface density or even to the model of the star-disc
interaction, we observed that the fraction of retrograde orbits captured
by the central mass in the course of their evolution depends on details of
both star-disc collisions and effects of the disc gravity. On the
other hand, our results show only a weak sensitivity on the total
mass of the accretion disc. This feature can be easily understood by
realizing that the disc mass parameter $\mu$ factorizes out
(in the first order of approximation) from the averaged potential
${\bar V}_{\rm d}$. As a consequence, the value of $\mu=10^{-3}$ taken
in our examples in Sec.~3 is not essential, and similar results
hold also for less massive discs.

Note: we have prepared a Java animation which illustrates long-term
evolution of stellar orbits in the two zones of the $(k,h)$-plane
(libration and oscillation in eccentricity); cf.\
``\hp\lb{2}karas/\lb{2}papers/\lb{2}discapplet.\lb{2}html''.
\medskip

We are grateful to Richard Stark and the unknown referee for very helpful 
suggestions concerning the presentation of our article. 
We acknowledge support from grants GA\,CR\,205/\lb{2}97/\lb{2}1165
and GA\,CR\,202/\lb{2}96/\lb{2}0206 in the Czech Republic.
V.~K. is grateful for kind hospitality of the International Centre
for Theoretical Physics and International School for Advanced Studies
in Trieste where this work was completed.


\appendix
\section{Gravitational field of a thin disc}
\label{appa}
In this section we describe a general method for evaluating
the gravitational potential, and its gradient, of an axisymmetric disc.
We introduce cylindrical coordinates $({R},z,\phi)$ with
origin in the centre of the disc and the plane $z=0$ coinciding with
the disc plane. The gravitational potential of the disc (outer radius
$b_\di$) evaluated at arbitrary position $({R},z)$ is given by (see,
e.g., Binney \& Tremaine 1987)

\beq
 V_\di\left({R},z\right) = -\,4G \int_0^{b_\di} \frac{\akpa({R^\prime})
  {R^\prime}}{B({R^\prime})}
  K\left[k\left({R^\prime}\right)\right]\,\myder{{R^\prime}} \, ,
  \label{21one}
\eeq
where
\beq
 k^2\left({R^\prime}\right) = \frac{4{R} {R^\prime}}{B^2({R^\prime})} \,,
 \label{21two}
\eeq
and
\beq
 B^2\left({R^\prime}\right) = z^2 + \left({R}+{R^\prime}\right)^2 \, .
 \label{21three}
\eeq
Here, $K(k)$ is the elliptic integral of the first kind. (The dependence
of functions $B({R^\prime})$ and $k({R^\prime})$ on coordinates ${R}$
and $z$ is not indicated explicitly.)

No apparent reduction of integral (\ref{21one}) is possible until the
radial distribution of the density $\akpa({R^\prime})$ is specified. An
important example which we will need below is a uniform disc,
$\akpa({R^\prime})\equiv\akpa_0$. The potential~(\ref{21one}) expressed
in terms of elliptic integrals $K$, $E$ and $\Pi$ reads, for
${R}<b_\di$, (Lass~\& Blitzer 1983)

\begin{eqnarray}
\lefteqn{V_\di({R},z)_{\mid\akpa=\akpa_0} \,\equiv\,
 V_\uni\left({R},z;\akpa_0\right)}
  \nonumber \\
 &=& 2G\akpa_0 \biggl\{\pi |z| - B\left(b_\di\right) E\left[k\left(
  b_\di\right)\right] - \frac{b_\di^2 - {R}^2}{B\left(
  b_\di\right)} K\left[k\left(b_\di\right)\right]
    \nonumber \\
 & & - \frac{b_\di-{R}}{b_\di+{R}}
 \frac{z^2}{B\left(b_\di\right)} \, \Pi\left[\alpha^2;
 k\left(b_\di\right)\right]\biggr\}
 \label{21four}
\end{eqnarray}
with
\beq
 \alpha^2 = \frac{4b_\di {R}}{(b_\di+{R})^2} \, . \label{21five}
\eeq
Formula (\ref{21four}) holds also in the region ${R}>b_\di$ provided
the first term in brackets of the right-hand side is suppressed. The
expression for the potential inside the disc $(z=0,$ ${R}<b_\di)$ can
be simplified by applying the Gauss transformation of elliptic functions
(Byrd~\& Friedman 1971). One finds that

\beq
 V_\uni({R},0;\akpa_0)=-4Gb_\di\akpa_0 E\left({R}/b_\di\right)
  \, , \label{21six}
\eeq
a more compact formula than the one given by Lass~\& Blitzer (1983).

The components of gravitational force are given by the gradient of
the potential (\ref{21one}). Direct algebraic manipulation results in

\begin{eqnarray}
 \frac{\partial V_\di}{\partial {R}} &=& -\frac{2G}{{R}}
  \int_0^{b_\di}
  \frac{\akpa({R^\prime}) {R^\prime}}{B({R^\prime})} \biggl
  \{E\left[k\left({R^\prime}\right)\right]
  \frac{{R^\prime}^2-{R}^2+z^2}{A^2\left({R^\prime}\right)} \nonumber \\
 & & -K\left[k\left({R^\prime}\right)\right]
  \biggl\}\,\myder{{R^\prime}}\,,  \label{21seven} \\
 \frac{\partial V_\di}{\partial z} &=& -4Gz \int_0^{b_\di}\frac{
  \akpa({R^\prime})\,
  {R^\prime}\,E\left[k\left({R^\prime}\right)\right]}{B({R^\prime})\,A^2
  \left({R^\prime}
  \right)}\,\myder{{R^\prime}}\,, \label{21eight}
\end{eqnarray}
where we denoted
\beq
 A^2\left({R^\prime}\right)
 = z^2 + \left({R}-{R^\prime}\right)^2 \, . \label{21nine}
\eeq
For a uniform disc one obtains
\begin{eqnarray}
 \frac{\partial V_\uni}{\partial {R}} &=& -\frac{2G\akpa_0}{{R}
  B\left(b_\di\right)} \Bigl\{B^2\left(b_\di\right) E\left[k\left(b_\di\right)
  \right] \nonumber \\
 & & -\left(z^2+b_\di^2+{R}^2
  \right) K\left[k\left(b_\di\right)\right] \Bigr\} \, , \label{21ten} \\
 \frac{\partial V_\uni}{\partial z} &=& 2G\akpa_0 \biggr\{\pm \pi -
  \frac{z}{B\left(b_\di\right)} \Bigl( K\left[k\left(b_\di\right)\right]
  \nonumber \\
 & & +\frac{b_\di-{R}}{b_\di+{R}} \, \Pi\left[\alpha^2\left(b_\di\right);
  k\left(b_\di\right) \right] \Bigr) \biggr\} \, . \label{21eleven}
\end{eqnarray}
The integrands in eqs. (\ref{21one}) and
(\ref{21seven})--(\ref{21eight}) diverge in the disc plane,
$z\rightarrow0,$ although the result of integration must be finite
because the potential is continuous across the disc. Taking into account
relation

\beq
 K(k)\propto\ln\left(1-k^2\right)  \label{21twelve}
\eeq
for $k\approx 1$, one concludes the divergence in the potential is
proportional $\ln z$. To get rid of numerical errors one conveniently
splits the integrals into two parts by setting $\akpa({R^\prime})\equiv
[\akpa({R^\prime})-\akpa({R})]+ \akpa({R})$. Then the potential is

\begin{eqnarray}
 V_\di\left({R},z\right)
 &=& -4G \int_0^{b_\di} \frac{\left[\akpa({R^\prime})-
  \akpa({R})\right] {R^\prime}}{B({R^\prime})} \,
  K\left[k\left({R^\prime}\right)\right]\,\myder{{R^\prime}}
  \nonumber \\
 & & +V_\uni\left[{R},z;\akpa({R}) \right]\,,
  \label{21thirteen}
\end{eqnarray}
where the second term corresponds to the potential of the disc with constant
density $\akpa({R})$ given by eq. (\ref{21four}). Now the integrand in
(\ref{21thirteen}) is well behaved. A sharp decrease of the integrand
near ${R^\prime}\approx{R},$ $z\approx 0$ can be treated by appropriate
numerical methods. The same approach can be applied successfully to
evaluate the components of force in eqs.\ (\ref{21seven})--(\ref{21eight}).

Previous formulae are easily generalized to the case when the inner edge
of the disc is at radius $a_\di\neq 0$. The lower integration limit in
(\ref{21one}) and (\ref{21seven})--(\ref{21eight}) is changed to
$a_\di$, and in the case of formulae (\ref{21four}) and
(\ref{21ten})--(\ref{21eleven}) for the uniform density disc one
employs a superposition of the field of a fictitious uniform disc with
radius $a_\di$ and formal density~$-\akpa_0$.

\label{lastpage}
\end{document}